\documentclass[10pt,letterpaper,twocolumn]{article} %% two column, final layout
\usepackage{ol2}
\usepackage[draft]{hyperref}
\usepackage{amsmath}

\begin{document}

\twocolumn[ %% activate for two-column option

\title{Photon correlations for colloidal nanocrystals and their clusters}

%% For REVTeX it is possible to automate superscript and e-mail callouts with the superscriptaddress option; see REVTeX4 documentation.

\author{O.~A.~Shcherbina$^{1}$, G.~A.~Shcherbina$^{2}$, M.~Manceau$^{3}$, S.~Vezzoli$^{3}$, L.~Carbone$^{4}$, M.~De Vittorio$^{4}$, A.~Bramati$^{3}$, E.\,Giacobino$^{3}$,  M.~V.~Chekhova$^{1,5,6,*}$, G.~Leuchs$^{5,6}$}

\address{
$^1$M.V.Lomonosov Moscow State University,
119992 GSP-2 Moscow, Russia \\
$^2$Moscow Institute of Physics and Technology,
141700 Dolgoprudny, Moscow Reg., Russia\\
$^3$Laboratoire Kastler Brossel, Universit\'e Pierre et Marie Curie,
Ecole Normale Sup\'erieure, CNRS, 4 place Jussieu, 75252 Paris Cedex 05, France\\
$^4$National Nanotechnology Laboratory - Istituto Nanoscienze CNR, Via Arnesano 16 - 73100 Lecce, Italy\\
$^5$Max-Planck Institute for the Science of Light, G.-Scharowsky Str 1/Bldg 24, 91058, Erlangen, Germany\\
$^6$Universit\"at Erlangen-N\"urnberg, Staudtstra{\ss}e 7/B2, 91058 Erlangen, Germany\\
$^*$Corresponding author: drquantum@hotmail.com
}

\begin{abstract}
Images of semiconductor `dot in rods' and their small clusters are studied by measuring the second-order correlation function with a spatially resolving ICCD camera. This measurement allows one to distinguish between a single dot and a cluster and, to a certain extent, to estimate the number of dots in a cluster. A more advanced measurement is proposed, based on higher-order correlations, enabling more accurate determination of the number of dots in a small cluster. Nonclassical features of the light emitted by such a cluster are analyzed.
\end{abstract}


 ] %% activate for two-column option

\noindent
Many quantum information protocols, such as quantum key distribution, linear optical quantum computation, and quantum metrology, require on-demand single-photon sources~\cite{Bouwmeester,single}. Among the sources available now in the laboratory, there are atoms~\cite{atoms}, molecules~\cite{molecules}, nitrogen vacancies in diamond~\cite{diamond}, and epitaxial semiconductor quantum dots~\cite{dots}. Colloidal quantum dots~\cite{colloidal}, despite their disadvantages such as blinking~\cite{blinking}, bleaching~\cite{bleaching}, and a noticeable probability of two-photon emission, are still one of the most promising types of single-photon emitters as they can be operated at room temperature and can be synthesized relatively easily. Especially promising are dot-in-rods (DR)~\cite{DR}, which have a higher probability of single-photon emission, show reduced blinking~\cite{Pisanello_th}, and emit single photons with a high degree of polarization~\cite{Pisanello}.

DRs can also merge into clusters. A cluster of DRs is not a single-photon emitter but it can be used for quantum information purposes since it still emits nonclassical light. In this work we perform measurements that allow one to distinguish between a single DR and a cluster. As a criterion, we use Glauber's second-order correlation function (CF) and the brightness of emission. Another method of distinguishing between a single DR and a cluster is defocused microscope technique~\cite{Pisanello}, based on the dipole-like angular distribution of the DR emission; however, this method only allows one to recognize clusters consisting of differently oriented DRs. As an extension of our correlation measurement, we propose the study of higher-order correlation functions. Such a measurement will allow one to resolve the number of DRs in a cluster and also observe nonclassical behavior of its emission.

For our experiment we used an intensified CCD (ICCD) camera. This device is equivalent to an array of single-photon detectors and can be used for spatially resolved measurements of Glauber's CFs. Alternatively, the pixels can be used for photon-number, rather than spatial, resolution, so that higher-order CFs can be measured. The ICCD camera can be used in either analog or photon-counting mode. In the photon-counting mode, used in our experiment, a certain level of the dimensionless readout signal $S$ (proportional to the integral electric charge acquired in a pixel) is chosen as a threshold, and any signal exceeding this value is interpreted as a single-photon event in a corresponding pixel. If the threshold $S_{th}$ is taken too high, the resulting quantum efficiency becomes low, which fortunately does not affect the normalized CF measured in our experiment~\cite{1986}. If the threshold is too low, the noise is increased.

\begin{figure}[h!]
\begin{center}
\includegraphics[width=0.8\columnwidth]{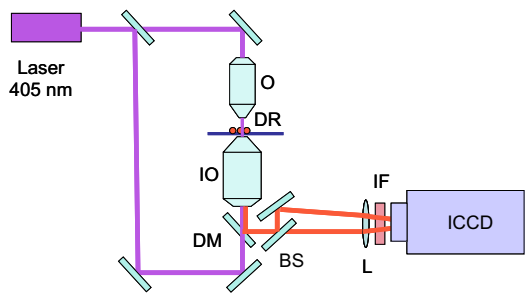}
\caption{The experimental setup.} \label{setup}
\end{center}
\end{figure}
The experimental setup is shown in Fig.~\ref{setup}. CdSe/CdS DRs with $4.6$ nm core diameter, $29$ nm shell length and $11$ nm shell width were dissolved in toluene with the concentration $10^{-13}$ mol/l and coated onto a glass substrate. After drying, the sample was placed over an NA$1.3$ oil immersion objective (IO) with the DRs on the side opposite to the objective. The excitation was performed by irradiating with cw diode laser light at $405$ nm. To slow down the bleaching of the DRs, the laser beam was modulated with a frequency of $30$ Hz and a pulse duration of $550\,\mu$s. There was a possibility of excitation through the same objective IO as well as through an NA0.75 objective (O) placed on top of the sample. This way a larger number of DRs could be excited. The excitation rate was chosen to be close to the saturation level.

\begin{figure}[h!]
\begin{center}
\includegraphics[width=0.9\columnwidth]{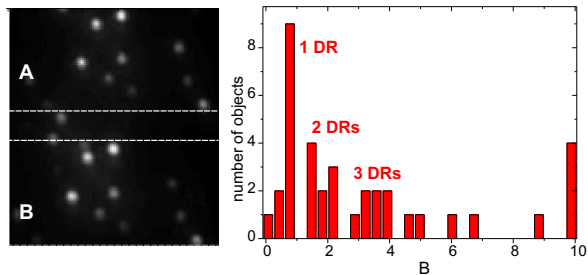}
\caption{Left: an image of several DRs and their clusters. Right: brightness distribution for the objects registered by the ICCD camera. The first three groups of bars are associated with single DRs and 2-DR and 3-DR clusters, respectively.} \label{DRs_and_hist}
\end{center}
\end{figure}

The emission from the DRs, centered at $630$ nm, was collected by the immersion objective and sent to the registration part of the system by means of a dichroic mirror (DM). Afterwards, the beam was split in two by a beamsplitter (BS), and both beams were directed to the Princeton Instruments PI-MAX3:1024i ICCD camera (ICCD) at a small angle. The path length difference between the beams was $20$ cm, corresponding to a time delay far below the smallest time scale in the experiment ($10$ ns). A lens (L) with the focal length $30$ cm was placed in front of the camera so that the photocathode was in its focal plane, producing two images of a group of DRs and their clusters (parts A and B of the image shown in Fig.~\ref{DRs_and_hist}, left). An interference filter with a width of $40$ nm and a central wavelength of $650$ nm was placed in front of the camera to reduce the contribution of stray light.

The ICCD camera was gated synchronously with the laser pulses, and the gate width could be varied between $10$ ns and $40$ ns. In order to optimize the single-photon detection, binning of pixels was performed (every group of 4x4 pixels was joined into a single `superpixel') and the readout threshold was chosen. For this, an area around a single DR was selected and the number of single-photon counts over this area (`the signal') was compared with the corresponding number for an identical empty area (`the noise'). The threshold value $S_{th}=685$ corresponded to the largest signal-to noise ratio. However, in experiment this value was slightly changed (within the range $685<S_{th}<700$) in order to minimize the error in the correlation function measurement. The resulting signal-to-noise ratio always exceeded 3.

For a measurement of the second-order CF, a set of $10^6$ frames was acquired. Because of the limited gate rate, the whole data acquisition took many hours. The unavoidable small displacements of the images during this time (due to small temperature variations and the mechanical vibrations) were taken into account by taking a long frame after every $10^4$  standard frames. These `control' frames allowed us to trace the displacement of the DR images and to make appropriate corrections to the coordinates of pixels chosen in each frame.

Five datasets were taken, with the gate times $10$, $15$, $20$, $30$, and $40$ ns. For each dataset, the mean number of single-photon events per `superpixel' per frame was less than $0.1$. The probability of having two photons taken for one was therefore negligible. Second-order CF at zero delay and displacement, $g^{(2)}(0,0)$, was calculated as
\begin{equation}
g^{(2)}(0,0)=\frac{\langle N_A N_B\rangle}{\langle N_A\rangle\langle N_B\rangle},
\label{g2}
\end{equation}
where the angular brackets denote averaging over all frames and $N_{A,B}$ are numbers of single-photon events for an area associated with a given object, either a DR or a cluster, in the fields $A,B$ ($N_{A,B}=0,1$). Besides, for each object seen in the image (Fig.~\ref{DRs_and_hist}, left), the dimensionless `brightness' $B$ was calculated taking into account the average number of counts $\langle N_{A,B}\rangle$ per gate, the gate time $T_g$, and the intensity $I$ of the excitation beam at the position of the object:
\begin{equation}
B\equiv \frac{\langle N_A\rangle }{T_g\tilde{I}}.
\label{B}
\end{equation}
Here, $\tilde{I}=\alpha I$ and $\alpha$ is a normalization factor, the same for all measurements, which has been chosen in such a way that most frequent objects have a value of $B$ close to $1$.
The distribution of the objects with respect to the brightness value $B$ is shown in Fig.~\ref{DRs_and_hist}, right. One can see that, despite the low total number of objects in the image, several groups can be seen: roughly, with $B<1.25$, $1.25<B<2.5$, and $2.5<B<4$. The comparison with the CF measurements below will show that these groups very likely relate to a single DR and to clusters of 2 and 3 DRs, respectively.
Due to the mechanical vibrations in the system as well as to the bleaching and blinking of the DRs, the brightness of each object changed during the data acquisition time. This was taken into account by normalizing the resulting CF $g^{(2)}(0,0)$ of each object to its value at a time delay of $T=30$ ms, $g^{(2)}(T,0)$. The latter was calculated by taking frames separated by $30$ ms, which is much larger than the lifetime. The values of $g^{(2)}(0,0)$ were also corrected taking into account the level of the noise, for which $g^{(2)}(0,0)=1$.

\begin{figure}[h!]
\begin{center}
\includegraphics[width=0.6\columnwidth]{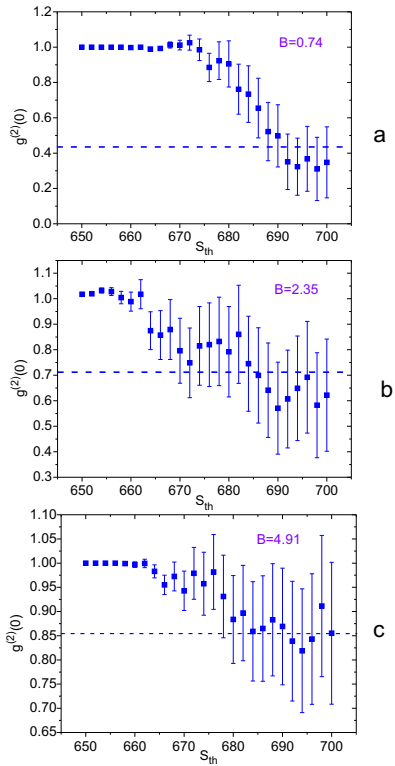}
\caption{The bunching parameter measured for individual objects in Fig.~\ref{DRs_and_hist} versus the threshold value: (a) a `dim object', $B=0.74$; (b) a `medium object', $B=2.35$; (c) a `bright object', $B=4.91$. The gate width is $10$ ns. Dashed lines: theoretical predictions for 1,2, and 4 dots.} \label{g2_th}
\end{center}
\end{figure}

\begin{figure}[h!]
\begin{center}
\includegraphics[width=0.6\columnwidth]{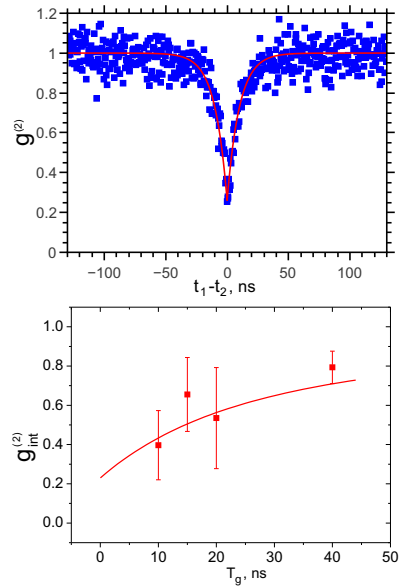}
\caption{Top: $g^{(2)}(t_1-t_2)$ dependence obtained by coincidence measurement with two avalanche photodiodes. Bottom: integrated bunching parameter $g^{(2)}_{int}$ averaged over all `dim' objects ($B<1.25$) as a function of the gate (integration) time $T_g$. Solid line: fit using (\ref{g2meas}).} \label{g2_T}
\end{center}
\end{figure}

The results of $g^{(2)}$ measurement with the gate time $T_g=10$ns are shown in Fig.~\ref{g2_th}. The measurement error is estimated from the number of registered single-photon events. For all objects, no anti-bunching is observed at low threshold values (Fig.~\ref{g2_th}a). This is because at low threshold, the noise is too high and its Poissonian statistics masks the anti-bunching. At higher threshold values, we see that an object that most probably is a single DR ($B=0.74$) manifests significant anti-bunching: $g^{(2)}=0.35\pm0.1$.
However, this value is relatively high because of the large gate width ($T_g=10$ ns).
 The effect of the averaging over $T_g$ can be described as follows. For cw excitation~\cite{Pisanello_th, Pisanello}, the CF depends on the times of the first and second photons registration $t_1,t_2$ as
\begin{equation}
g^{(2)}(t_1,t_2)=1-(1-p)\exp[-k|t_1-t_2|],
\label{g2t}
\end{equation}
where $k$ is the decay rate and $p$ the probability of two-photon emission. In order to determine these parameters, $g^{(2)}(t_1-t_2)$ was measured in a different setup, with a single DR excited at the saturation level and the emission registered by two avalanche photodiodes followed by a coincidence circuit. The results (Fig.~\ref{g2_T}, top) show that $k=0.1$ ns$^{-1}$ and $p=0.22$ if the noise contribution into $g^{(2)}(0)$ is eliminated. Integration of Eq.~(\ref{g2t}) in $t_1,t_2$ within the limits from $0$ to $T_g$ yields
\begin{equation}
g^{(2)}_{int}(T_g)=1-(1-p)\{\frac{2}{kT_g}+\frac{2}{(kT_g)^2}(\exp[-kT_g]-1)\}.
\label{g2meas}
\end{equation}
From this expression, the expected value of the bunching parameter measured for a single DR with the gate time $10$ ns is $0.43$ (dashed line in Fig.~\ref{g2_th}), which is in agreement with the measured value.
For a brighter object ($B=2.35$), which we associate with a cluster of 2 DRs, the value of $g^{(2)}$ measured at optimal threshold values is $0.65\pm0.1$ (Fig.~\ref{g2_th}b). Theoretically, the existence of two single-photon emitters in a cluster can be taken into account by using the general formula describing $g^{(2)}$ in the presence of $m$ independent contributions (modes)~\cite{single-mode},
\begin{equation}
g^{(2)}_m=1+\frac{g^{(2)}_1-1}{m},
\label{m}
\end{equation}
where $g^{(2)}_1$ is the value for a single emitter. For $m=2$ and $g^{(2)}_1=0.43$, we get $g^{(2)}_2=0.72$, in agreement with the data in Fig.~\ref{g2_th}b. Finally, Fig.~\ref{g2_th}c shows the bunching parameter for an object with $B=4.91$, for which we assume $m=4$. This gives the theoretical prediction $g^{(2)}_4=0.86$, again in good agreement with the measurement ($0.85\pm0.1$). The conclusion follows that our hypothesis about the numbers of DRs in clusters, based on their brightness, agrees very well with the bunching parameter measurements. This gives a good indication for the relevance of our method to evaluate the number of nonclassical emitters in a cluster.

Figure~\ref{g2_T} (bottom) shows the results of averaging the measured bunching parameter over all objects for which $B<1.25$ and which are therefore identified as single DRs. The threshold was chosen to be $693$. We see that the value of the bunching parameter grows with the gate width. The dependence is well fit with Eq.~(\ref{g2meas}). The errors take into account both the number of single-photon events and the spread of the data obtained for different DRs. In total, $10$ DRs contributed into the plot, but only 3 of them, for $T_g=10$ns, 2 for $T_g=15$ns, 2 for $T_g=20$ns, and 3 for $T_g=40$ns, as the statistical error was too large for the rest.

It is also interesting to consider the measurement of higher-order CFs for a DR cluster.
%Although with our current setup it is not possible, using faster gating (for instance, with a picosecond excitation source) would make it feasible.
Indeed, for a cluster of $N$ DRs, the $N$th-order CF will be nonzero, $g^{(N)}(0,0)\ne0$ while the next-order one will show the analog of anti-bunching, $g^{(N+1)}(0,0)\approx0$. This is because a cluster of $N$ single-photon emitters cannot emit more than $N$ photons within the lifetime. From such measurements, one can get more information about the number of DRs in a cluster. Note that here, it is not required that all DRs emit photons into the same radiation mode. Even for a product state of $N$ photons simultaneously emitted into different modes, the nonclassicality condition~\cite{DNK}
\begin{equation}
g^{(N-1)}g^{(N+1)}<[g^{(N)}]^2
\label{nonclas1}
\end{equation}
will be satisfied.

Moreover, if all DRs in the cluster emit photons into the same mode, even with small (but equal) probabilities, another nonclassical feature can be observed. Indeed, the cluster emits then a Fock state
\begin{equation}
|N\rangle=\frac{1}{\sqrt{N!}}(a^{\dagger})^N|\hbox{vac}\rangle.
\label{N}
\end{equation}
In the presence of losses leading to a finite detection efficiency $\eta$, one can account for the latter using the beamsplitter model, with the transmission coefficient $t\equiv\sqrt{\eta}$ and the reflection coefficient $r\equiv\sqrt{1-\eta}$. Then the light state becomes
\begin{equation}
|\Psi\rangle=\sum_{i=0}^N\frac{\sqrt{N!}}{\sqrt{i!(N-i)!}}t^ir^{N-i}|i\rangle|N-i\rangle,
\label{state}
\end{equation}
and the probability to register $k$ photons is
\begin{equation}
p_k=\frac{N!}{k!(N-k)!}\eta^k(1-\eta)^{N-k}.
\label{prob}
\end{equation}
Regardless of the detection efficiency, this probability distribution satisfies another nonclassicality condition~\cite{DNK},
\begin{equation}
\frac{k+1}{k}\frac{p_{k+1}p_{k-1}}{p_k^2}<1,
\label{nonclass}
\end{equation}
since for the state (\ref{state})
\begin{equation}
\frac{k+1}{k}\frac{p_{k+1}p_{k-1}}{p_k^2}=\frac{N-k}{N-k+1}.
\label{satisf}
\end{equation}
(Note that for a Poissonian state, both (\ref{satisf}) and (\ref{nonclas1}) will give 1.) The nonclassical behavior will be especially noticeable for $k$ close to $N$.

In conclusion, we have observed spatially resolved images of several single-photon emitters (`dot-in-rods') and their clusters. Using a single-photon photodetector array (ICCD camera), we were able to measure the bunching parameter for each individual object and to distinguish a single emitter from a cluster of such. By assuming that the brightness of an object scales as the number of DRs in it, we identified two-dot, three-dot, and four-dot clusters. This assumption is confirmed by the results of the correlation function measurement. Finally, we propose the measurement of higher-order correlations as a method for a better determination of the number of single-photon emitters in a cluster.

This work was supported by ERA-Net.RUS (project Nanoquint) and by the Russian Foundation for Basic Research, grant 12-02-00965. O.~A.~S. acknowledges support from the Dynasty Foundation. We are grateful to M.~Sondermann and V.~Salakhutdinov for helpful discussions.


\begin{thebibliography}{99}
\bibitem{Bouwmeester} D.~Bouwmeester, A.~Ekert, and A.~Zeilinger, \textit{The Physics of Quantum Information} (Springer-Verlag, Berlin, 2000).

\bibitem{single}S.~Buckley, K.~Rivoire and J.~Vuckovic, ``Engineered quantum dot single-photon sources", Rep. Prog. Phys. \textbf{75}, 126503 (2012).

\bibitem{atoms}M.~Hijlkema, B.~Weber, H.~P.~Sprecht, S.~C.~Webster, A.~Kuhn, and G.~Rempe, ``A
single-photon server with just one atom", Nature Phys. \textbf{3}, 253 (2007).

\bibitem{molecules}F.~Treussart,  R.~Allaume, V.~Le Floch, L.~Xiao, J.~Courty, and J.~Roch, ``Direct measurement of the photon statistics of a triggered single photon source", PRL \textbf{89}, 93601 (2002).

\bibitem{diamond}A.~Gruber, A.~Dr\"abenstedt, C.~Tietz, L.~Fleury, J.~Wrachtrup, and C.~von Borczyskowski, ``Scanning Confocal Optical Microscopy and Magnetic Resonance on Single Defect Centers", Science \textbf{276}, 1997 (2012).

\bibitem{dots}C.~Santori, M.~Pelton, G.~Solomon, Y.~Dale, and Y.~Yamamoto, ``Triggered single photons from a quantum dot", PRL \textbf{86}, 1502 (2001).

\bibitem{colloidal}P.~Michler, A.~Imamoglu, M.~D.~Mason, P.~J.~Carson, G.~F.~Strouse, and S.~K.~Buratto, ``Quantum correlation among photons from a single quantum dot at room temperature", Nature  \textbf{406}, 968 (2000).

\bibitem{blinking}A.~Efros and M.~Rosen, ``Random telegraph signal in the photoluminescence intensity
of a single quantum dot", PRL \textbf{78}, 1110 (1997).

\bibitem{bleaching}W.~G.~J.~H.~M. van Sark, P.~L.~T.~M. Frederix, D.~J.~Van den Heuvel, and H.~C.~Gerritsen, ``Photooxidation and Photobleaching of Single CdSe/ZnS Quantum Dots Probed by Room-Temperature Time-Resolved Spectroscopy", J. Phys. Chem. B  \textbf{105}, 8281 (2001).

\bibitem{DR}L.~Carbone, C.~Nobile, M.~D.~Giorgi, F.~D.~Sala, G.~Morello, P.~Pompa, M.~Hytch, E.~Snoeck, A.~Fiore, I.~R.~Franchini, M.~Nadasan, A.~F.~Silvestre, L.~Chiodo, S.~Kudera, R.~Cingolani, R.~Krahne, and L.~Manna, ``Synthesis and micrometer-scale assembly of colloidal cdse/cds nanorods prepared by a seeded
growth approach", Nano Letters \textbf{7}, 2942 (2007).

\bibitem{Pisanello_th}
F. Pisanello, ``Single Photon Sources Based on Colloidal Nanocrystals and Two Photon Polariton Laser", Ph.D. thesis (2013).

\bibitem{Pisanello}
F. Pisanello, L.~ Martiradonna, G.~Leménager, P.~Spinicelli, F.~Fiore, L.~Manna,
J.~Hermier, R.~Cingolani, E.~Giacobino, M.~De Vittorio, and A.~Bramati, ``Room
temperature-dipolelike single photon source with a colloidal dot-in-rod", Appl. Phys. Lett. \textbf{96}, 033101 (2010).

\bibitem{1986}
G.~Leuchs, \textit{Photon Statistics, Antibunching and Squeezed States. In 'Frontiers of Non-Equilibrium Statistical Physics', G.T. Moore und M.O. Scully (Eds.)}, p, 329 (Plenum Press, NY, London, 1986).

\bibitem{single-mode}
T.~Sh.~Iskhakov, A.~Perez, K.~Yu.~Spasibko, M.~V.~Chekhova, and G.~Leuchs, ``Statistics of single-mode bright squeezed vacuum", Optics Lett. \textbf{37}, 1919 (2012).

\bibitem{DNK}
D.~N.~Klyshko, ``The nonclassical light", Physics-Uspekhi \textbf{39}, 573 (1996).
\end{thebibliography}
\end{document}